\documentclass[preprint,12pt,authoryear]{elsarticle}

\usepackage{amssymb}
\usepackage{hyperref}
\usepackage{graphicx}
\usepackage{subfig}
\journal{Network Analysis}

\begin{document}

\begin{frontmatter}



%%\title{Link prediction in Foursquare dataset\tnoteref{label1}}
%%\author{Urban Marovt, Rok Fortuna\corref{cor1}\fnref{label2}}

\title{Link prediction in Foursquare network}


\author{Rok Fortuna, Urban Marovt}

\address{University of Ljubljana \\ 
Faculty of Computer and Information Science \\
rok.fortuna@cloud.si, urban.marovt@cloud.si}

\begin{abstract}
Foursquare is an online social network and can be represented with a bipartite network of users and venues. A user-venue pair is connected if a user has checked-in at that venue. In the case of Foursquare, network analysis techniques can be used to enhance the user experience. One such technique is link prediction, which can be used to build a personalized recommendation system of venues. Recommendation systems in bipartite networks are very often designed using the global ranking method and collaborative filtering. A less known method- network based inference is also a feasible choice for link prediction in bipartite networks and sometimes performs better than the previous two. In this paper we test these techniques on the Foursquare network. The best technique proves to be the network based inference. We also show that taking into account the available metadata can be beneficial.

\end{abstract}


\end{frontmatter}


\section{Introduction}
Bipartite networks are one of the most common types of graphs in real world problems: drug-target networks \citep{yildirim2007drug}, author-paper \citep{zhou2007bipartite} networks, user-product networks \citep{zhou2007bipartite} etc. They consist of nodes of two types and nodes are only connected to those of the different type. Foursquare social network \citep{chorley2011checking} is a network of users checking in at venues and can be modeled using a bipartite network. A check-in is therefore an edge with some metadata if viewed from the network analysis point of view. This metadata includes the time of the check-in and the rating a user has given to that venue. The metadata can be used to improve the performance of the network analysis techniques such as link prediction.

Network analysis and machine learning methods can be used to enhance the user experience by providing users a recommended list of venues \citep{savage2012m}. Most network analysis methods are designed for unipartite graphs which have only one type of nodes. However we can transform bipartite graphs to weighted unipartite graphs using projections and use the same network analysis methods for them. In the case of Foursquare social network, such a projection yields a weighted unipartite graph of users or venues. By analyzing these projections we can obtain additional information for our link prediction analysis (e.g. community detection).

The goal of this study is to exploit the Foursquare network structure and the metadata available to build a feasible link prediction method. The method should perform better than the classic recommendation system methods (global ranking method, collaborative filtering etc.) and possibly be used to enhance the Foursquare recommendation system.
\label{}


\section{Related work}
\label{}

Since we have chosen Foursquare social network, we do not have simple graph to analyze, but the structure of the network is bipartite (two-mode). In \citep{latapy2008basic} authors describe how to deal with basic bipartite network analysis and introduce some properties we can only find in bipartite networks. 

In (\cite{allali2013internal, zhou2007bipartite, everett2013dual}) authors suggest to analyze bipartite networks by projecting them to one-mode networks and analyzing the projections. Authors claim that by projecting bipartite networks the loss of information is insignificant, therefore classic (one-mode) network analysis can be applied. In (\cite{allali2013internal}) authors describe an approach, that predicts edges that do not have any influence on the later projection to an one-mode network.

\cite{zhou2007bipartite} discuss how to compute the similarity between nodes in a projected network. Since our network is not only bipartite, but also contains multilinks, we can extract more information by applying weights to the edges. We can compute similarities between users which incorporate the weights and use that information to enhance our prediction. In (\citep{liu2015stability}) authors discuss some approaches and properties that can help constructing such similarity measures.

In (\cite{scellato2011exploiting}) authors discuss the problems of link prediction in location-based social networks. They introduce some important parameters (e.g. latitude and longitude of a check-in) that are taken into account in their link prediction method.

\section{Methods}
\label{methods}

Link prediction is a network analysis problem where we want to predict whether a link between two nodes that are not linked yet will occur \citep{allali2013internal}. In bipartite graphs we are only predicting the links between different types of nodes since only these kind of links exist. 

In the case of Foursquare, its recommendation system recommends most suitable venues that a certain user has not yet checked in at. The success of such a system is measured by how useful these recommendations actually were (how it affected the user's future check-ins). This problem is obviously analogous to the problem of bipartite link prediction. 

\subsection{Methods for link prediction}
\label{methods_used}
We often perform link prediction in bipartite graphs using one-mode projections (\cite{allali2013internal, everett2013dual}). In other words, we build an unipartite network and preserve as much information as possible from the bipartite structure. In the case of Foursquare we achieve that by building a graph of solely users or solely venues. Commonly used methods for link prediction include global ranking method (GRM), collaborative filtering (CF) and network based inference (NBI) \citep{zhou2007bipartite}.

GRM simply ranks the nodes (in our case venues) based on their degrees. In our case links including higher ranked venues are predicted. 

\label{collaborative_filtering}
CF is based on node similarities (similarities between the same type of nodes). In our case they are computed for each pair of users. Two users are more similar if they have both checked in at more common venues. There are many node similarity measures (\cite{liu2015stability}). In this paper, we use the Adamic-Adar index \citep{feng2012link}. Based on user similarities for a certain user we then obtain the score for every venue by computing the sum of similarities of other users that have checked in at that venue. At the end we normalize the score with the sum of all similarities for the user. For every user, links to venues with higher scores are predicted. With CF we also take into account the structure of the bipartite network (in contrast to, GRM which produces global scores). 

\label{network_based_inference}
NBI is based on resource allocation. For every user the following procedure is executed. We assign some initial resource to the venues. This step depends on our domain knowledge and other metadata we have available. The most basic way is assigning a unit weight to those venues the user has already checked in at. We then perform the two step resource flow. In the first step each venue equally distributes its resource to the users that have checked in at it. The second step is analogous as the users equally distribute their resource back to the venues. A user is more likely to connect to higher ranked venues (venues with more resource), therefore links to the venues with higher score are predicted. NBI also takes into account the network structure (as does CF).

We also evaluate the very basic assortativity method. It predicts that users who mostly check-in at lower degree venues more likely to check-in at another lower degree venue and vice versa. 

\subsection{Evaluating link prediction method performance}
\label{AUC}
Each of the methods described in the previous section gives a user-venue score for each of the possible user-venue pairs. For the performance evaluation measure we use the area under the ROC (Receiver Operating Characteristics) curve - AUC (\cite{huang2005using}). 

As input it takes two sets of nodes (in our case user-venue pairs). The first set contains user-venue pairs which are actually linked while the other set contains ones that are not. We first build a set of user-venue pairs that are not connected. After that we build a set of user-venue pairs that are connected. In each step after adding a connected user-venue pair to our sample, we remove all of the remaining edges between the chosen user-venue pair. 

We compute the AUC by sampling one user-venue pair from each of the sets at random and comparing their scores according to the chosen link prediction method. A link prediction method is successful if more connected user-venue pairs have higher scores than the ones that are not. AUC basically measures the probability that a pair of connected nodes will have a higher link prediction value than a pair of not connected nodes. Its values range from $0.5$ to $1$. With $0.5$ being the score of picking link prediction values for every user-venue pair at random.

\subsubsection{Sampling}
\label{set_sampling}
Both sets, the one that contains user-venue pairs that are actually connected and the second one that contains ones that are not, are constructed at random. First we select the fraction of nodes we would like to sample. Then we randomly pick pairs that are not connected and pairs that are connected. We also remove all multilinks (excess check-ins) between each selected pair. 

As we are provided with exact times of the check-ins it is reasonable to also construct samples based on time. Besides sampling links at random we also sample the user-venue pairs based on a selected time period. For every chosen pair we first remove all of the check-ins between selected user and venue, then compute the pair score and reconstruct the original network.

\section{Results}

\label{basic_results}

The Foursquare dataset was adopted from (\cite{yang2014modeling}). The network contains $227426$ check-ins from $1083$ users at $38333$ venues. 

\begin{figure}[!ht] \centering
	\subfloat[Users degree distribution]{\includegraphics[width=0.4325\textwidth]{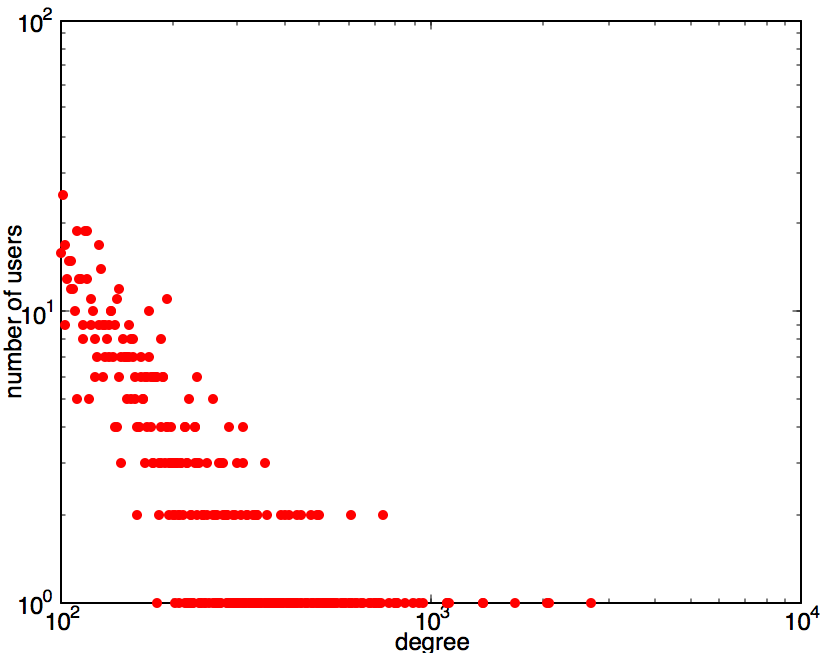}}
	\subfloat[Venues degree distribution]{\includegraphics[width=0.4325\textwidth]{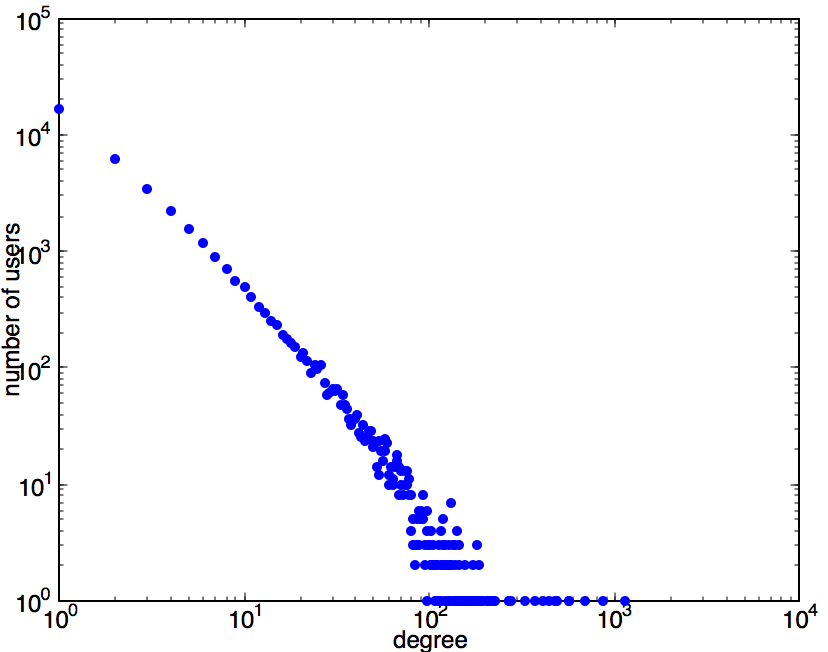}}
	\caption{{\bf Degree distributions.}}
	\label{fig:degreeDistr}
\end{figure}

Figure \ref{fig:degreeDistr} contains plotted degree distributions of users and venues. We can see that there are no users with low degrees. All users have degrees larger than $100$. On the other hand we can see that there are a lot of venues with low degrees. We do not want to predict links of users to very low degree venues as predicting such a link would almost definitely be faulty. That is why filtering the graph by removing low degree venues results in a better network for analysis. 

By analyzing the graph structure we can also notice there are many venues where only one user (or maybe very few users) are checking-in at. It is also beneficial to remove these kind of venues as they are most definitely user specific (e.g. home, work etc.). 

Because of the reasons stated above, we remove venues with degrees lower than $20$ and those for which only one user accounts for $90\%$ of the check-ins. We obtain a graph of $1083$ users, $1267$ venues and $62478$ check-ins. Approximately $60\%$ of all of the edges are removed as a consequence of removing low degree venues and additional $10\%$ due to venues having very few different users checking-in at them.

\subsection{Sampling the network}

We sample our data in two ways, at random and based on a time period. When sampling at random we build samples in advance and then run our algorithms. This is computationally much faster than sampling links one by one and running the analysis. Due to method time complexity it is very hard to incrementally test methods that are based on user similarities (e.g. CF). 

\renewcommand{\figurename}{Figure}
\begin{figure}[ht!] \centering
	\fbox{\includegraphics[width=0.55\textwidth]{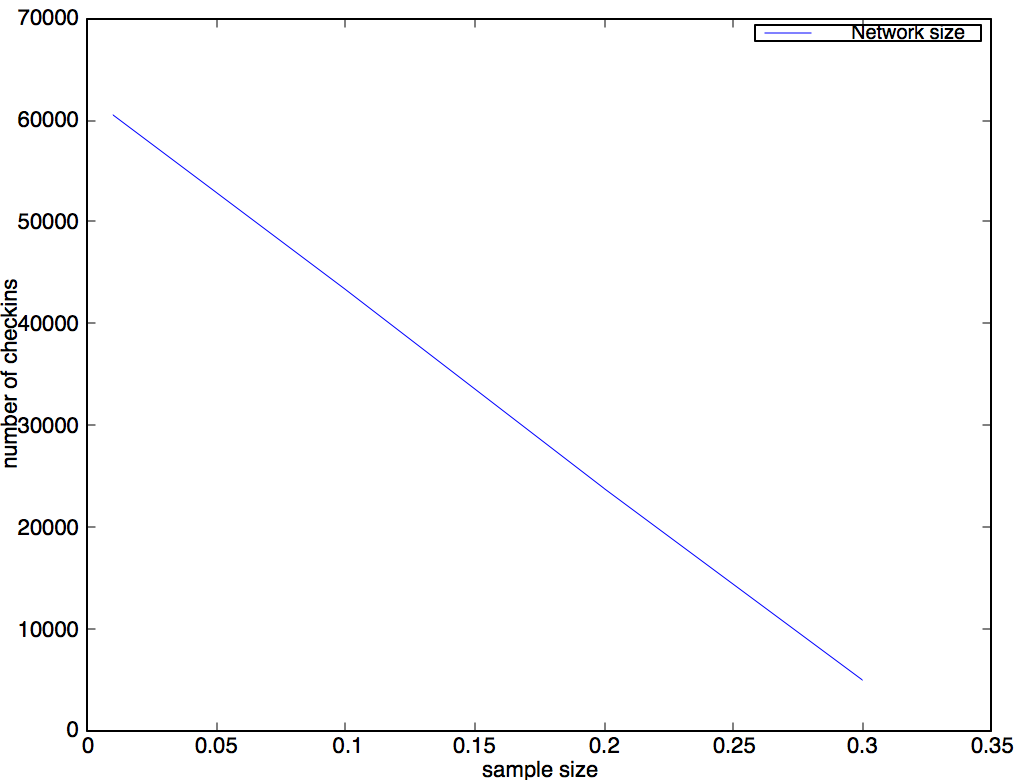}}
	\caption{Number of check-ins for various random sample sizes.}
	\label{network_size}
\end{figure}

On the other hand a problem that arises with batch sampling is that the network looses its structure very rapidly. Large parts of the network are removed due to removing multilinks. Figure \ref{network_size} shows the remaining number of check-ins after the sampling. With a sample size of $30\%$ we are left with less than $10\%$ of the check-ins. This implies that the network completely looses it's structure and the results for such samples are unreliable. 

\clearpage
\subsection{Metadata importance}
\label{metadata_importance}
\begin{figure}[ht!] \centering
	\fbox{\includegraphics[width=0.6\textwidth]{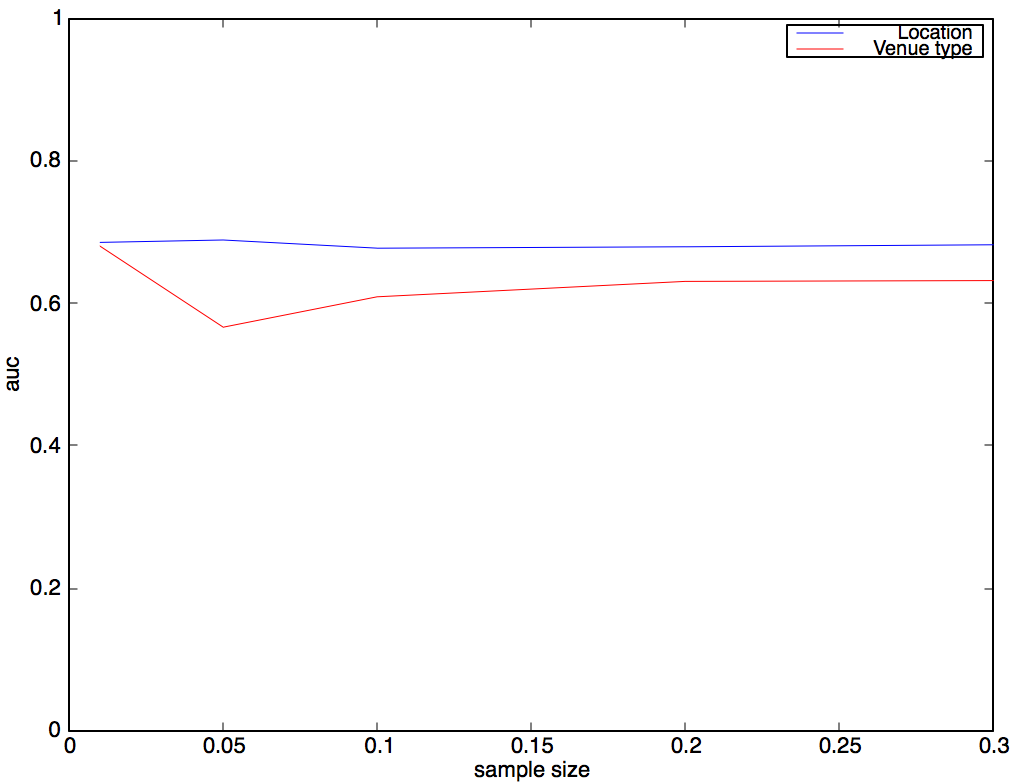}}
	\caption{AUC results for metadata importance.}
	\label{meta_method_results}
\end{figure}

Figure \ref{meta_method_results} includes the AUC scores which represent the importance of metadata. "Location" represents the meaningfulness of the distance between the average location of user's check-ins and the venue. "Venue type" for a given user-venue pair from the sample is computed as the fraction of user's check-ins with the same venue type as the one in the sample. The higher these scores are, the more related these meta attributes are to the occurrence of the links.

\renewcommand{\figurename}{Figure}

From the figure \ref{meta_method_results} it is clear that there is some correlation between the occurrence of a link and the provided meta attributes. The results show that the metadata can be used as a potential improvement to link prediction methods.

\subsection{Link prediction methods}

\renewcommand{\figurename}{Figure}
\begin{figure}[ht!] \centering
	\fbox{\includegraphics[width=0.6\textwidth]{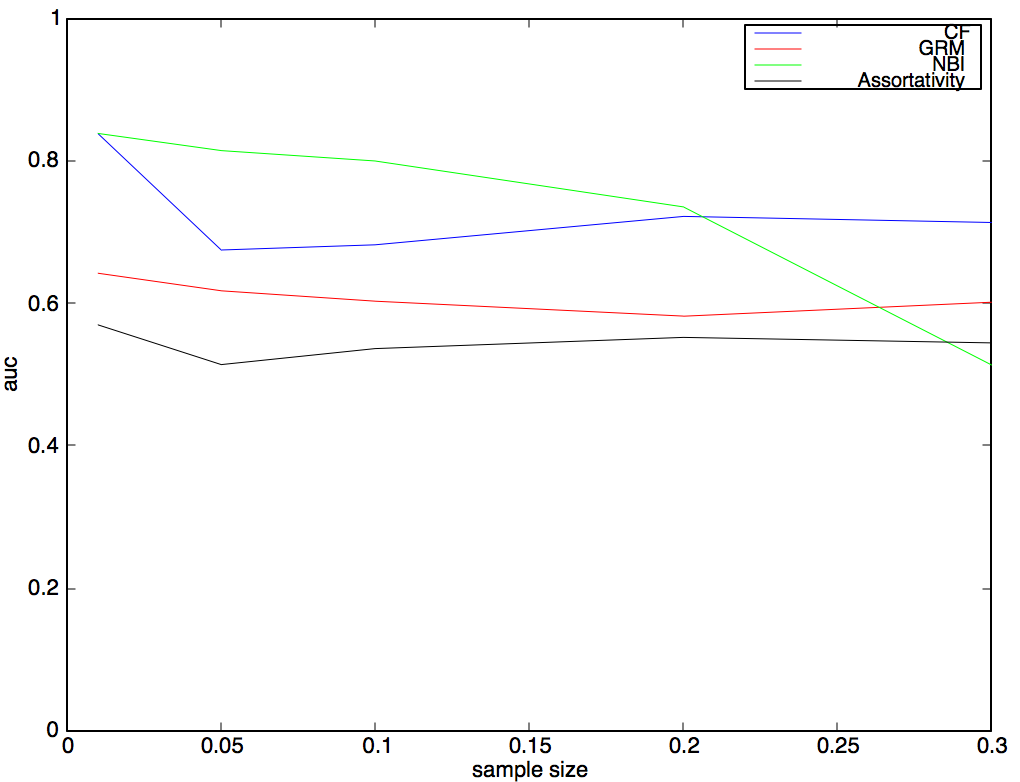}}
	\caption{AUC results for various sample sizes.}
	\label{basic_method_results}
\end{figure}

Figure \ref{basic_method_results} displays the AUC scores of the evaluated methods. The methods used include collaborative filtering (CF), global ranking method (GRM), network based inference (NBI) and assortativity method (see section \ref{methods_used}).\\

We can see that NBI by far outperforms other methods, except at sample size of $30\%$. However, the result of the $30\%$ sample is meaningless, since the network looses most of its structure. From the chart we can clearly see that assortativity method and GRM perform about the same regardless of the sample size. That is due to being mostly independent of the network structure (they are based on venue degrees). CF performs quite well and clearly beats the two degree-based methods. The obtained results imply that network structure holds a lot more information than just analyzing the venues by themselves.

\subsection{Derived methods} 
\label{customized_nbi_methods}
From the previous section we can conclude that network structure needs to be used in order to provide good results. But in section \ref{metadata_importance} we also reason that these methods could benefit from the metadata provided. In this section we take the best performing method - NBI and modify it in various ways. The resource allocation step in the NBI method can be modified by using the metadata provided by the dataset to distribute resources accordingly (instead of simply giving a unit resource to each of the venues a user has checked in at). 

\begin{figure}[ht!] \centering
	\fbox{\includegraphics[width=0.6\textwidth]{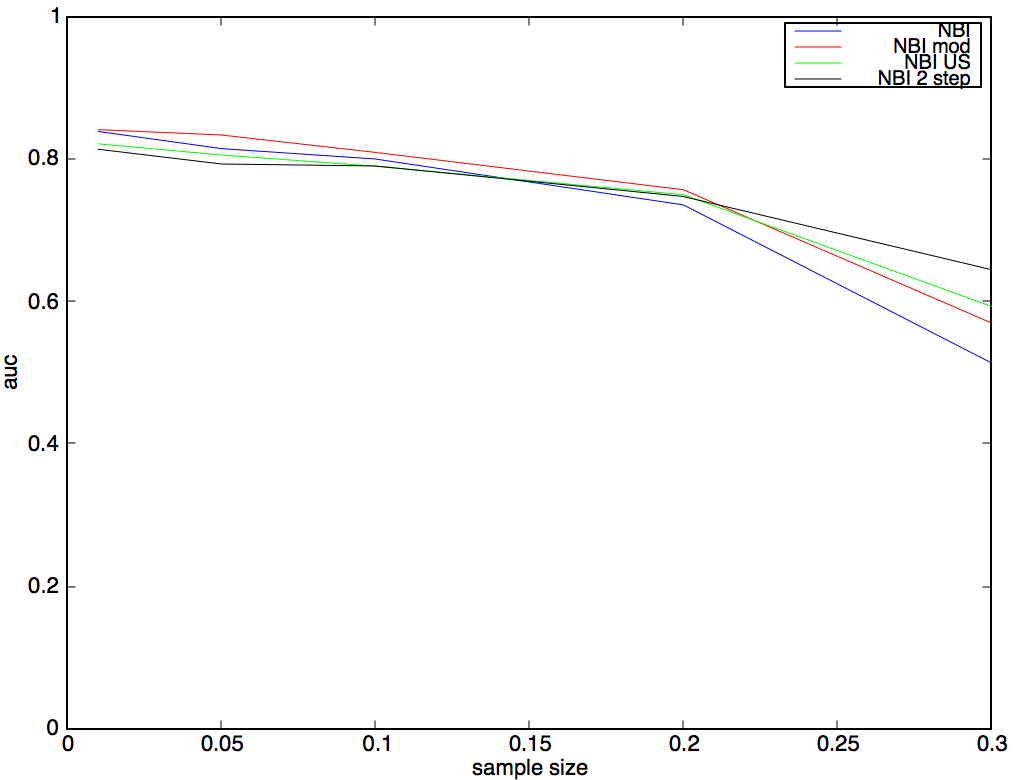}}
	\caption{AUC results for the derived NBI methods.}
	\label{method_results_customized}
\end{figure}

Figure \ref{method_results_customized} includes the following plots. "NBI mod" represents the NBI method which takes into the account the global degree of a venue, the venue type (see section \ref{metadata_importance}) and the location. "NBI US" is very similar to "NBI mod" but also allocates some resources to users based on the user similarity to the user in the sample (similarity is in accordance with the one in CF, see section \ref{collaborative_filtering}). "NBI 2 step" denotes the basic NBI algorithm with two step flow (the resource flows four times instead of two). 

\renewcommand{\figurename}{Figure}

We can see that the modified NBI comes out on top. Clearly the metadata can be used to provide additional useful information but its contribution is not very significant. The other two methods (the one using user similarities and two-step NBI) perform worse than the basic NBI. Regarding the NBI using user similarities, we can state that they do not provide any additional information to the NBI method. As for the two-step NBI we can conclude that the additional spread of resources (the two step spreading) does not give enough resource to other venues important to the user being analyzed. It turns out the more iterations we make the worse AUC we get. That is simply explainable as the resource flows through entire network and venue relevance fades.

\subsection{Time sampling}
\label{time_sampling}
\begin{figure}[ht!] \centering
	\fbox{\includegraphics[width=0.6\textwidth]{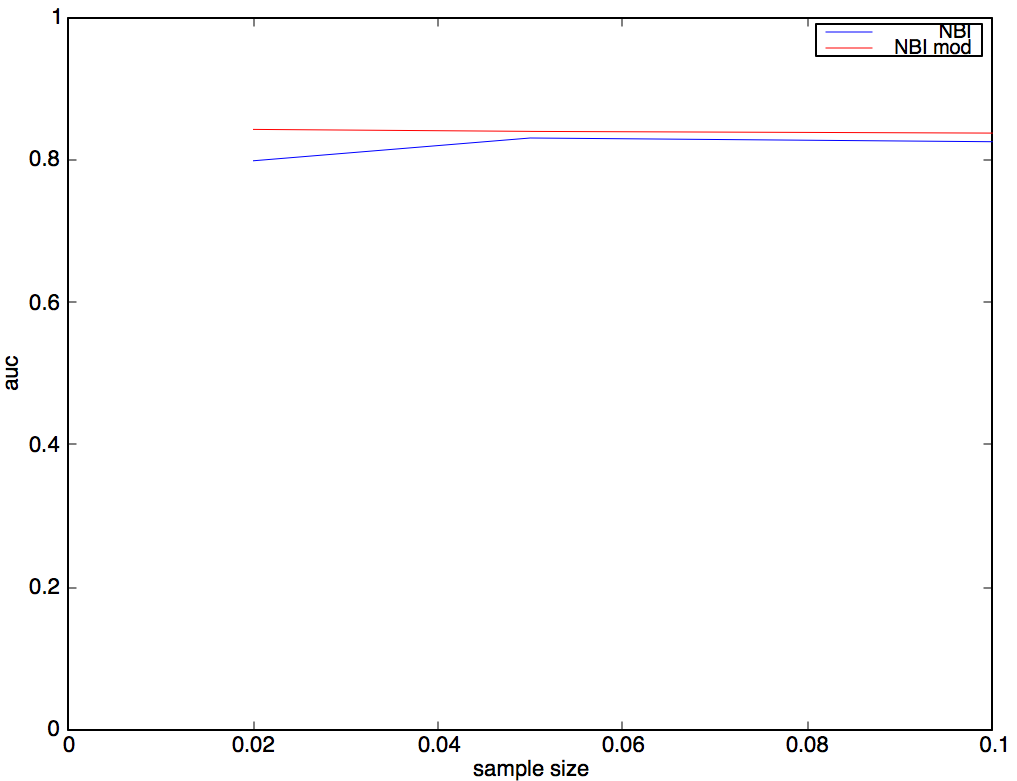}}
	\caption{AUC results for the time samples of various sizes.}
	\label{method_results_time}
\end{figure}

In the previous sections we use the random sampling technique to obtain the AUC sample while in this section we use the the sampling based on time (see section \ref{set_sampling}). Figure \ref{method_results_time} displays the results two different methods on a time sample. The "NBI" denotes the basic NBI method without any modifications (see section \ref{network_based_inference}). "NBI time mod" takes into account how trendy the pair's venue is. The trend of a particular venue is computed based on the fraction of check-ins that occurred close to the selected period of time. If many users checked in recently (close to the selected period), then the venue is trendy.  

\renewcommand{\figurename}{Figure}

It is evident that we can exploit the time meta attribute and outperform the basic NBI. The difference is not very significant but still noticeable. The Foursquare data has a timespan of 10 months. The data includes all four seasons of the year which results in a season-based trendiness of venues. By taking trendiness into account we can enhance the user experience of a Foursquare recommendation system, by suggesting the season trendy venues.
\clearpage
\section{Conclusion}
In this paper we benchmark the well known bipartite link prediction methods on a Foursquare network. The network itself contains a lot of noise, since there are many venues with low degrees, and many venues having very few users checking in at them. For a feasible analysis, it is necessary to filter the network and to sample the network in such a way that it preserves the structure (see section \ref{set_sampling}). 

The methods we benchmark include GRM, CF, basic assortativity and NBI (see section \ref{methods}). We compare the success of these methods using the AUC measure (see section \ref{AUC}). We analyze two types of methods: global methods based on venue degrees (GRM and basic assortativity), and methods that take into account the structure of the network (CF and NBI). The second type of methods greatly outperforms the first one. This implies that personalized link prediction systems can be enhanced by taking into account the network structure. 

The overall winner method is the NBI which outperforms CF by roughly $10\%$ on a sample of size $10\%$. As the Foursquare network offers additional metadata we also incorporate it into the NBI method to obtain better prediction scores. The attributes that are beneficial to the NBI are the global degree of a venue, venue type and the location (see section \ref{customized_nbi_methods}). Other modifications of the NBI (e.g. two-step flow and using user similarities) are also benchmarked, but do not outperform the metadata-modified NBI.

As the Foursquare network is a time based network, we are also provided with the times of check-ins. In this paper we also introduce a modified version of NBI method which also relies on the trendiness of the venues (see section \ref{time_sampling}). This method is benchmarked using a time based sample built only by sampling edges in a certain timespan. It outperforms the classic NBI method, which proves that there is some correlation between the trendiness and link occurrence. This implies that venue trendiness is seasonal and can be incorporated in production recommendation systems.

\pagebreak
\bibliographystyle{elsarticle-harv} 
\bibliography{biblio}




%%\bibitem[ ()]{}

%%\end{thebibliography}
\end{document}